\documentstyle[12pt,epsf]{article}
\begin{document}

\thispagestyle{empty}
\begin{flushright}
SLAC-PUB-7575\\
TUM-HEP-280/97\\
hep-ph/9707243\\
July 1997
\end{flushright}
\vspace*{2cm}
\centerline{\Large\bf Two-Loop Large-\mbox{\boldmath $m_t$} 
             Electroweak Corrections to
             $K\to\pi\nu\bar\nu$}
\centerline{\Large\bf for Arbitrary Higgs Boson Mass
\footnote{
Work supported by the U.S. Department of Energy under contract
DE-AC03-76SF00515, by the German Bundesministerium 
f\"ur Bildung und Forschung under contract 06 TM 874 and
by the DFG project Li 519/2-2.}}

\vspace*{1.5cm}
\centerline{{\sc Gerhard Buchalla$^1$} 
and {\sc Andrzej J. Buras$^2$}}
\bigskip
\centerline{\sl $^1$Stanford Linear Accelerator Center}
\centerline{\sl Stanford University, Stanford, CA 94309, U.S.A.}
\vskip0.6truecm
\centerline{\sl $^2$Physik Department, Technische Universit\"at
                M\"unchen}
\centerline{\sl D-85748 Garching, Germany}

\vspace*{1.5cm}
\centerline{\bf Abstract}
\vspace*{0.2cm}
\noindent 
We consider for the first time the leading large top mass corrections,
arising at higher order in electroweak interactions,
to the rare decays $K\to\pi\nu\bar\nu$ and the related modes
$B\to X_s\nu\bar\nu$ and $B\to l^+l^-$.
Higher order effects of similar type have previously been calculated
in the large-$m_t$ limit for key observables of precision
electroweak physics at $Z$-factories. Here we obtain the corresponding
corrections of order ${\cal O}(G^2_F m^4_t)$ at the amplitude level for 
short-distance dominated rare meson decays. This allows us to quantify
the importance of higher order electroweak effects for these
processes, which can be reliably computed and have very small 
uncertainties from strong interactions.
Simultaneously it becomes possible to remove, to some extent,
ambiguities in the definition of electroweak parameters describing
the strength of FCNC interactions.
The corrections we discuss are at the level of a few percent.
\vspace*{0.5cm}
\noindent 

\vspace*{1.2cm}
\noindent
PACS numbers: 12.15.Lk, 13.20.Eb, 13.20.He

\vfill

\newpage
\pagenumbering{arabic}

\section{Introduction}
\label{intro}

In the Standard Model flavor-changing neutral current (FCNC)
interactions are generated at one-loop order. They give rise to
neutral meson mixing, CP violation and rare decays, which therefore 
provide excellent opportunities to study flavor dynamics. A class of
rare decay modes, including $K\to\pi\nu\bar\nu$, 
$B\to X_{s,d}\nu\bar\nu$ and $B_{s,d}\to l^+l^-$, has long been 
recognized to be particularly interesting in this respect. Since there 
are no contributions from virtual photons in these cases, the GIM
cancellation pattern is powerlike
($\sim m^2_i/M^2_W$, $i=u, c, t$, for $m_i\ll M_W$),
resulting in a strong suppression of potential long distance effects.
The processes are dominated by short distances, related to the
heavy particles ($W$, top, charm) in the loop, and can be reliably
calculated.
\\
The low-energy effective Hamiltonian for $K\to \pi\nu\bar\nu$
to lowest order in electroweak interactions can be written as
\begin{equation}\label{heff}
{\cal H}_{eff}=\frac{G_F}{\sqrt{2}}
\frac{\alpha}{2\pi\sin^2\Theta_W}\left(\lambda_t X_0(x_t)+
\lambda_c X_0(x_c)\right)(\bar sd)_{V-A}(\bar\nu_l\nu_l)_{V-A}+ h.c.
\end{equation}
where $\lambda_i=V^*_{is}V_{id}$ and $x_i=m^2_i/M^2_W$. Here the
lepton mass dependence (only important for the charm contribution
in the case of the $\tau$-lepton) has been neglected for simplicity.
The one-loop function is given by \cite{IL}
\begin{equation}\label{x0x}
X_0(x)=\frac{x}{8}\left[\frac{x+2}{x-1}+\frac{3x-6}{(x-1)^2}
\ln x\right]
\end{equation}
Only the top quark contribution is relevant for the CP violating
neutral mode $K_L\to\pi^0\nu\bar\nu$. For $K^+\to\pi^+\nu\bar\nu$
the charm sector contributes typically $40\%$ of the branching ratio
and is therefore not negligible, though still somewhat smaller than the
top contribution.

Eq. (\ref{heff}) provides a reasonable approximation as a
basis for calculating $K\to\pi\nu\bar\nu$. For $K^+\to\pi^+\nu\bar\nu$
the $\tau$-lepton mass effect \cite{IL} and leading logarithmic
QCD corrections \cite{NVZS,EH,DDG} are relevant in the charm sector
and have been known for some time.
\\
Over the years important refinements have been added in the 
theoretical treatment
of $K\to\pi\nu\bar\nu$. Long-distance contributions were estimated
quantitatively and could be shown to be essentially negligible,
as expected \cite{RS,HL,LW,FAJ}. The hadronic matrix elements
$\langle\pi|(\bar sd)_V|K\rangle$ can be extracted from the leading
semileptonic decay $K^+\to\pi^0e^+\nu$ using isospin symmetry.
Corrections due to isospin breaking from quark masses and 
electromagnetism have been computed in \cite{MP}.
Finally, the complete next-to-leading order QCD corrections are
known \cite{BB1,BB2,BB3}. The NLO result eliminates the dominant
uncertainties of the leading order predictions, improving the
precision of the theoretical calculation.
\\
All these developments have led to a fairly advanced and quantitative
understanding of, and good control over theoretical uncertainties.
They are at the level of $5\%$ for $K^+\to\pi^+\nu\bar\nu$, dominated
by the charm sector, and even considerably smaller (below $2\%$)
for $K_L\to\pi^0\nu\bar\nu$, where the charm contribution is absent.
Correspondingly the prospects for precision tests of Standard Model
flavor physics are quite promising \cite{BB6} (for recent discussions
of new physics possibilities see e.g. \cite{GN,BUR}). 
An ongoing search for $K^+\to\pi^+\nu\bar\nu$ has set a branching
ratio limit of $2.4\cdot 10^{-9}$ \cite{SA} and is approaching the 
Standard Model range at $\sim 10^{-10}$. The current published upper 
limit on $B(K_L\to\pi^0\nu\bar\nu)$ is $5.8\cdot 10^{-5}$ \cite{WEA}.
It is particularly encouraging
that various efforts are now under way to make the challenging
experiments possible that are needed for precise measurements of
$K^+\to\pi^+\nu\bar\nu$ \cite{CCTR} and $K_L\to\pi^0\nu\bar\nu$
\cite{AGS,KAMI,ISS}.

In this situation it is interesting to carefully consider 
presumably small effects that have so far always been neglected.
An example are electroweak radiative corrections of higher order, which
are expected to be reasonably small and are probably not the first issue
one would worry about in the context of rare decays. However, given the
high level of precision already obtained in the theory of 
$K\to\pi\nu\bar\nu$, a more quantitative estimate of these corrections is
certainly worth pursuing.
Moreover, non-decoupling effects, due to electroweak symmetry breaking,
grow with $m_t$ and could in principle be sizable.
To our knowledge, these higher oder electroweak corrections have not been
studied previously for FCNC rare decays. On the other hand such effects
have been calculated, to leading order in large $m_t$, for
precision electroweak physics at $Z$-factories
\cite{BBCCV,FTJ,DGS}. In this context one should stress that all existing
analyses of rare decays have intrinsic theoretical uncertainties
related to the definition of electroweak parameters. In particular:
\begin{itemize}
\item
There is an ambiguity in the value of $\sin^2\Theta_W$ entering the
rare decay branching ratio formulas. The various possible
definitions of this quantity differ by electroweak radiative corrections
that amount to several percent. The related uncertainty can only be
removed by considering higher order electroweak effects.
\item
An ambiguity further exists in whether the pole mass or the 
$\overline{MS}$ mass of the top quark should be used. 
With respect to QCD interactions this uncertainty has been eliminated
through the calculation of ${\cal O}(\alpha_s)$ corrections.
However, the ambiguity
is not only due to QCD but also due to electroweak effects,
which in view of large $m_t$ are not fully negligible.
\item
Next there are scale ambiguities related to the top quark Yukawa
coupling caused by the Higgs-top Yukawa interaction.
\item
Finally it is of interest to see the impact of the neutral Higgs boson
on FCNC processes.
\end{itemize}
Many of these issues have been discussed in the context of electroweak
precision tests, in particular in \cite{KS}, but have not been considered
in connection with rare decays. 
An exception are higher order corrections of
purely electromagnetic origin and the ambiguity between the fine
structure constant $\alpha=1/137$ and $\alpha(M_Z)=1/129$. The dominant
effects of this type have already been taken into account previously.
They are not related to large top quark Yukawa interactions and therefore
not our major concern in the present context. We will however briefly
address this topic further later on. 

It is the purpose of this paper to derive the higher order electroweak
effects, in the limit of large top quark mass, that correct the leading
Inami-Lim function for $K\to\pi\nu\bar\nu$. The explicit expressions
obtained will enable us to quantify the impact of these corrections.
The corresponding effects will also be discussed for 
$B\to X_s\nu\bar\nu$ and $B\to l^+l^-$.

\section{Leading Large-$m_t$ Corrections to $K_L\to\pi^0\nu\bar\nu$}
\label{klpn}

For definiteness we will focus our discussion first on
$K_L\to\pi^0\nu\bar\nu$ and generalize to the remaining cases 
at the end of this section.
\\
Large non-decoupling top quark effects from electroweak loops are
of the form $G_F m^2_t$ to leading order in the large-$m_t$ limit.
In the following we shall work through order $G^2_F m^4_t$, 
corresponding to two-loop electroweak effects. Such corrections modify
in particular the $Z$-boson--fermion coupling into an effective vertex
$V_{\bar ffZ}$ and one may write \cite{FTJ}
\begin{equation}\label{vffz}
V_{\bar ffZ}=-i(\sqrt{2}\varrho G_F)^{1/2}\frac{M_Z}{2}\gamma^\mu
\left((1+\tau_b)2 T_{3f}(1-\gamma_5)-4 Q_f\kappa\sin^2\Theta_W\right)
\end{equation}
Here $G_F$, $M_Z$ and $\alpha=1/137$ are taken to be the basic electroweak
parameters; $\sin^2\Theta_W\equiv 1-M^2_W/M^2_Z$ can be expressed in
terms of these three quantities. $T_{3f}$ and $Q_f$ denote the third
component of weak isospin and the charge of the fermion $f$,
respectively. $\varrho$ and $\kappa$ are universal, propagator-type
corrections. $\tau_b$ is a non-universal vertex correction, which depends
on $f$ through the top quark CKM couplings.
Denoting
\begin{equation}\label{xit}
\xi_t=\frac{G_F m^2_t}{8\sqrt{2}\pi^2}
\end{equation}
one has in the above mentioned approximation \cite{FTJ}
\begin{equation}\label{rhoxi}
\varrho=1+\Delta\varrho=1+3 \xi_t+ {\cal O}(\xi^2_t)
\end{equation}
\begin{equation}\label{tauxi}
\tau_b=-2 \xi_t\left(1+\tau^{(2)}_b\xi_t+ {\cal O}(\xi^2_t)\right)
\end{equation}
The two-loop function $\tau^{(2)}_b$ depends on both the top quark mass 
$m_t$ and the Higgs-boson mass $m_H$, and reads \cite{BBCCV,FTJ}
\begin{eqnarray}\label{taub2}
\tau^{(2)}_b &=& 9-\frac{13}{4}a-2a^2-\frac{a}{4}(19+6a)\ln a-
\frac{a^2}{4}(7-6a)\ln^2 a-\left(\frac{1}{4}+\frac{7}{2}a^2-3a^3\right)
\frac{\pi^2}{6} + \\
&+& \left(\frac{a}{2}-2\right)\sqrt{a} g(a)+(a-1)^2
\left(4a-\frac{7}{4}\right)L_2(1-a)-\left(a^3-\frac{33}{4}a^2+18a-7\right)
f(a) \nonumber
\end{eqnarray}
where
\begin{equation}\label{al2}
a=\frac{m^2_H}{m^2_t}\qquad\qquad L_2(1-a)=\int^a_1 dt\frac{\ln t}{1-t}
\end{equation}
\begin{equation}\label{gadef}
g(a)=\left\{ \begin{array}{ll}
              2\sqrt{4-a} \arccos\sqrt{a/4} & 
                          \mbox{for $0\leq a\leq 4$} \\
              \sqrt{a-4}\ln\frac{1-\sqrt{1-4/a}}{1+\sqrt{1-4/a}} &
                          \mbox{for $a\geq 4$}
             \end{array}
     \right.
\end{equation}
\begin{equation}\label{fadef}
f(a)=\int^1_0dt\left[ L_2(1-r(t,a))+\frac{r(t,a)}{r(t,a)-1}\ln r(t,a)
\right], \qquad  r(t,a)=\frac{1+(a-1)t}{t(1-t)}
\end{equation}
The expression in (\ref{taub2}) corresponds to the pole definition
of the top quark mass.
\\
Expression (\ref{vffz}) may be generalized to the case of the
FCNC vertex $V_{\bar sdZ}$ by introducing $\lambda_i=V^*_{is}V_{id}$,
summing over $i=u,c,t$, using CKM unitarity and noting that $\tau_b$
has to be set to zero for $i=u,c$. Additive universal contributions
drop out and one obtains
\begin{equation}\label{vsdz}
V_{\bar sdZ}=-i(\sqrt{2}\varrho G_F)^{1/2}\frac{M_Z}{2}\lambda_t
(-\tau_b)\gamma^\mu(1-\gamma_5)
\end{equation}
Note in particular that the tree level part of $V_{\bar ffZ}$ is
canceled through the GIM mechanism and $V_{\bar sdZ}$ is,
like $\tau_b$, a pure loop effect.
The coupling of $Z$ to neutrinos can also be read off from $V_{\bar ffZ}$
and is
\begin{equation}\label{vnnz}
V_{\bar\nu\nu Z}=-i(\sqrt{2}\varrho G_F)^{1/2}\frac{M_Z}{2}
\gamma^\mu(1-\gamma_5)
\end{equation}
Combining (\ref{vsdz}), (\ref{vnnz}) and (\ref{xit})--(\ref{tauxi}), an
effective Hamiltonian, valid to first and second order in
$G_F m^2_t$, can be constructed for $K_L\to\pi^0\nu\bar\nu$
\begin{equation}\label{hfcnc}
{\cal H}_{eff,FCNC}=\frac{G^2_F m^2_t}{16\pi^2}
\left(1+(3+\tau^{(2)}_b)\xi_t\right)\lambda_t 
(\bar sd)_{V-A}(\bar\nu\nu)_{V-A}+ h.c.
\end{equation}
(\ref{hfcnc}) coincides with (\ref{heff}) in the large top mass limit
and to leading order in electroweak interactions. The corresponding
effective Hamiltonian for the charged current process 
$K^+\to\pi^0e^+\nu$, useful for normalizing $K_L\to\pi^0\nu\bar\nu$,
is given by
\begin{equation}\label{hcc}
{\cal H}_{eff,CC}=\frac{G_F}{\sqrt{2}}V^*_{us}
(\bar su)_{V-A}(\bar\nu e)_{V-A}
\end{equation}
{}From (\ref{hfcnc}) and (\ref{hcc}) it is straightforward to obtain
($\lambda\equiv V_{us}$)
\begin{equation}\label{bkl1}
\frac{B(K_L\to\pi^0\nu\bar\nu)}{B(K^+\to\pi^0 e\nu)}=3
\frac{\tau_{K_L}}{\tau_{K^+}}\frac{G^2_F m^4_t}{64\pi^4}
\left[ 1+2(3+\tau^{(2)}_b)\xi_t\right]
\left(\frac{\mbox{Im}\lambda_t}{\lambda}\right)^2
\end{equation}
where we have summed over neutrino flavors.

\begin{figure}[t]
   \vspace{0cm}
   \epsfxsize=5cm
   \centerline{\epsffile{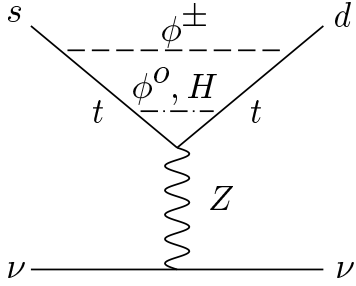}}
   \vspace*{-0cm}
\caption{\label{kpnwfig1} Typical diagram contributing at 
  ${\cal O}(G^2_F m^4_t)$ to the $K\to\pi\nu\bar\nu$ amplitude.
  Here $\phi^\pm$, $\phi^0$ are Higgs-ghosts and $H$ is the physical
  Higgs-particle. The complete set of diagrams can be found in
  \cite{FTJ}.}
\end{figure}

We remark that in \cite{BBCCV,FTJ} the effective vertex (\ref{vffz})
has been derived in the limit where all external momenta are negligible
in comparison with $m_t$ and $m_H$. Therefore the result is applicable
to both electroweak observables at the $Z$-boson resonance, considered
in \cite{BBCCV,FTJ}, as well as to low energy effective Hamiltonians
for rare meson decays that we are interested in here.
Note also that in the large-$m_t$ limit only $Z$-penguin but no box
diagrams contribute to $K\to\pi\nu\bar\nu$. A typical two-loop
electroweak diagram relevant for the ${\cal O}(G^2_F m^4_t)$ correction
to the decay amplitude is shown in Fig. \ref{kpnwfig1}. 

Next we recall that, within the approximation we need for our
purposes, one has \cite{JEG}
\begin{equation}\label{gfr}
G_F=\frac{\pi\bar\alpha}{\sqrt{2}M^2_W\sin^2\Theta_W}
\left(1-\frac{\cos^2\Theta_W}{\sin^2\Theta_W}\Delta\varrho\right)
\end{equation}
where $\sin^2\Theta_W\equiv1-M^2_W/M^2_Z$ (on-shell definition),
$\bar\alpha\equiv\alpha(M_Z)=1/129$ and $\Delta\varrho=3\xi_t$.
Using (\ref{gfr}) in (\ref{hfcnc}) one has
\begin{equation}\label{hfc2}
{\cal H}_{eff,FCNC}=\frac{G_F}{\sqrt{2}}
\frac{\bar\alpha}{2\pi\sin^2\Theta_W}\lambda_t\frac{x_t}{8}
\left(1+\left(\tau^{(2)}_b+6-\frac{3}{\sin^2\Theta_W}\right)\xi_t\right)
(\bar sd)_{V-A}(\bar\nu\nu)_{V-A} + h.c.
\end{equation}
and (\ref{bkl1}) becomes
\begin{equation}\label{bkl2}
\frac{B(K_L\to\pi^0\nu\bar\nu)}{B(K^+\to\pi^0 e\nu)}=3
\frac{\tau_{K_L}}{\tau_{K^+}}
\frac{\bar\alpha^2}{2\pi^2\sin^4\Theta_W}
\left[ 1+\left(2\tau^{(2)}_b+12-\frac{6}{\sin^2\Theta_W}\right)\xi_t\right]
\left(\frac{x_t}{8}\right)^2
\left(\frac{\mbox{Im}\lambda_t}{\lambda}\right)^2
\end{equation}
This expression is useful since it contains the leading electroweak
coupling constants in the form conventionally chosen in analyzing
$K_L\to\pi^0\nu\bar\nu$. The various forms in which the electroweak
parameters may be written are all equivalent at lowest order, where
for instance $\sqrt{2}G_F M^2_W\sin^2\Theta_W=\pi\bar\alpha$. These
expressions differ by terms of order ${\cal O}(\xi_t)$ (see (\ref{gfr})).
Consequently, the explicit ${\cal O}(\xi_t)$ correction will be different
for different choices of electroweak couplings, while physical quantities
remain unchanged (compare (\ref{bkl1}) and (\ref{bkl2})).
\\
{}From the above derivation it is clear that the appropriate QED
coupling entering (\ref{hfc2}) and (\ref{bkl2}) is
$\bar\alpha=\alpha(M_Z)=1/129$ and not the usual fine structure
constant $\alpha=1/137$. These two quantities differ by logarithmic
terms $\sim\alpha\ln M_Z/m_f$.
On the other hand, the ratio in (\ref{bkl2}) does in fact receive a
logarithmic QED correction $\sim\alpha\ln M_Z/m_K$ not displayed in
this equation. It is due to the differences in the QED renormalization
between the neutral current and the charged current transitions forming
the ratio (\ref{bkl2}). This effect, which in principle is of similar 
nature as the difference between $\alpha$ and $\bar\alpha$, has been
discussed in \cite{MP} in the context of isospin breaking corrections.
We will not include this correction here, with the understanding that it
is part of the known isospin breaking effects \cite{MP} to be taken 
into account in a complete analysis of $K_L\to\pi^0\nu\bar\nu$.

If we use the Hamiltonian in the form of (\ref{hfc2}), the leading 
large-$m_t$ electroweak correction to $K_L\to\pi^0\nu\bar\nu$
may be written as a factor
\begin{equation}\label{rxew}
r_{X,EW}=1+\frac{x_t}{4 X_0(x_t)}
\left(\tau^{(2)}_b+6-\frac{3}{\sin^2\Theta_W}\right)\xi_t
\end{equation}
multiplying the leading order $K_L\to\pi^0\nu\bar\nu$ branching ratio.
Here we have generalized the lowest order top-mass dependence
$x_t/8$ to the complete function $X_0(x_t)$ (\ref{x0x}).
Only the leading large-$m_t$ terms have been kept for the
electroweak correction, as the full mass dependence to this order
is still unknown.
Equivalently we may express the electroweak effects as a correction to
the lowest order Inami-Lim function $X_0(x_t)$, which then becomes
\begin{equation}\label{xxpole}
X_0(x_t)+\frac{x_t}{8}\left(\tau^{(2)}_b+6-\frac{3}{\sin^2\Theta_W}
\right)\xi_t
\end{equation}
This modification likewise affects the top contribution to
$K^+\to\pi^+\nu\bar\nu$. However, because of the sizable charm
contribution that dominates the theoretical uncertainties in this case,
electroweak corrections are less relevant here. 

The same factor $r_{X,EW}$ applies also to the rare decay
$B\to X_s\nu\bar\nu$, whose branching fraction is to lowest order
given by \cite{BBL}
\begin{equation}\label{bbsnn}
\frac{B(B\to X_s\nu\bar\nu)}{B(B\to X_c e\nu)}=
\frac{3\bar\alpha^2}{4\pi^2\sin^4\Theta_W}
\left|\frac{V_{ts}}{V_{cb}}\right|^2\frac{X^2_0(x_t)}{f(m_c/m_b)}
\end{equation}
with the $b\to c e\nu$ phase space factor
$f(z)=1-8z^2+8z^6-z^8-24 z^4 \ln z$.

A closely related decay is $B_s\to l^+l^-$. The effective Hamiltonian
for this case is similar to (\ref{hfc2}) and can be obtained by
replacing $V^*_{ts}V_{td} (\bar sd)_{V-A}(\bar\nu\nu)_{V-A}$
$\to$ $-V^*_{tb}V_{ts} (\bar bs)_{V-A}(\bar ll)_{V-A}$. The leading
order $m_t$-dependence $x_t/8$ generalizes here to the function
\begin{equation}\label{y0x}
Y_0(x)=\frac{x}{8}\left[\frac{x-4}{x-1}+\frac{3 x\ln x}{(x-1)^2}\right]
\end{equation}
The branching ratio \cite{BBL}
\begin{equation}\label{bbll}
B(B_s\to l^+l^-)=\tau(B_s)\frac{G^2_F}{\pi}
\left(\frac{\bar\alpha}{4\pi\sin^2\Theta_W}\right)^2 F^2_{B_s} m^2_l
m_{B_s}\sqrt{1-4\frac{m^2_l}{m^2_{B_s}}}|V^*_{tb}V_{ts}|^2 Y^2_0(x_t)
\end{equation}
is then modified through large-$m_t$ electroweak effects by a factor
\begin{equation}\label{ryew}
r_{Y,EW}=1+\frac{x_t}{4 Y_0(x_t)}
\left(\tau^{(2)}_b+6-\frac{3}{\sin^2\Theta_W}\right)\xi_t
\end{equation}
This corresponds to a correction of the Inami-Lim function in
(\ref{y0x}), which gets replaced by
\begin{equation}\label{yxpole}
Y_0(x_t)+\frac{x_t}{8}\left(\tau^{(2)}_b+6-\frac{3}{\sin^2\Theta_W}
\right)\xi_t
\end{equation}
The functions $X_0$ and $Y_0$ differ only by box-diagram contributions.
Because these are vanishing in the large-$m_t$ limit, the correction
terms in (\ref{xxpole}) and (\ref{yxpole}) are identical.

\section{Numerical Results and Discussion}

In the following section we will present numerical results 
for the electroweak corrections and further discuss various aspects
of the analysis. To this purpose we specify first the relevant
input parameters. We will use
\begin{equation}\label{gfmw}
G_F=1.16639\cdot 10^{-5}\, GeV^{-2} \qquad M_W=80.34\, GeV
\end{equation}
\begin{equation}\label{swmt}
\sin^2\Theta_W\equiv 1-M^2_W/M^2_Z=0.2238 \qquad m_t=167\, GeV
\end{equation}
For the $W$-boson mass $M_W$ we take the central value of the
Standard Model prediction \cite{PDG}. The logarithmic dependence of
$M_W$ on the Higgs boson mass 
($\sim\xi_t\ln(m_H/M_W)\cdot M^2_W/m^2_t$)
can consistently be neglected within our approximation and is also
numerically small.
The value of $m_t$ in (\ref{swmt}) corresponds to the $\overline{MS}$
definition with respect to QCD corrections. It differs from the
value of the QCD pole mass $m_{t,pole(QCD)}=175\, GeV$ by about
$8\, GeV$. The $\overline{MS}$ definition is an appropriate choice
in the analysis of QCD effects, which have been discussed elsewhere
\cite{BB1,BB2}. On the other hand, the top quark mass $m_t$ will
here be understood to refer to the pole mass definition with respect to
electroweak effects. This is the choice that has been used in obtaining
the electroweak corrections in the previous section.

Numerical values for the correction factors $r_{X,EW}$ and $r_{Y,EW}$
are displayed in Table \ref{rxytab} for various values of the
Higgs-boson mass $m_H$.
\begin{table}
\begin{center}
\begin{tabular}{|c||c|c|c|c|c|c|}\hline
$m_H/GeV$ & 60 & 150 & 300 & 450 & 600 & 1000 \\
\hline
\hline
$r_{X,EW}-1$ & $-0.91\%$ & $-1.20\%$ & $-1.27\%$ & 
               $-1.17\%$ & $-1.02\%$ & $-0.58\%$ \\
\hline
$r_{Y,EW}-1$ & $-1.41\%$ & $-1.87\%$ & $-1.97\%$ & 
               $-1.82\%$ & $-1.59\%$ & $-0.91\%$ \\
\hline
\end{tabular}
\end{center}
\caption[]{The leading large-$m_t$ electroweak correction
factors $r_{X,EW}$ and $r_{Y,EW}$, as defined in 
(\protect\ref{rxew}) and (\protect\ref{ryew}), respectively,
for various values of the Higgs-boson mass $m_H$.
They multiply the branching fractions of
$K_L\to\pi^0\nu\bar\nu$, $B\to X_s\nu\bar\nu$ ($r_{X,EW}$) and 
$B\to l^+l^-$ ($r_{Y,EW}$).
\label{rxytab}}
\end{table}
The leading large-$m_t$ corrections shown there are moderate and amount
to typically $-1\%$. The largest effect is obtained for $m_H$ around
$170-340\, GeV$, where the (positive-valued) function
$\tau^{(2)}_b$ has a minimum. The corrections $r_{Y,EW}-1$ are larger 
than $r_{X,EW}-1$ by a factor of $1.56$ for $m_t=167\, GeV$ and
independent of $m_H$. They can reach values up to $-2\%$.
\\
We recall that these corrections depend on the form in which the
leading electroweak coupling constants are expressed. The factors
$r_{X,EW}$ and $r_{Y,EW}$ refer to the choice of
$\bar\alpha^2/\sin^4\Theta_W$ as used in (\ref{bkl2}), (\ref{bbsnn})
and (\ref{bbll}), with the on-shell Weinberg angle.
If instead one were to use the coupling expressed
in terms of the effective weak mixing angle
$\sin^2\hat\Theta(M_Z)=0.23$ \cite{PDG}, where\footnote{Note that the
approximate relations in (\ref{gfr}) and (\ref{st2hat}) hold quite
accurately for realistic values of the parameters.}
\begin{equation}\label{st2hat}
\sin^2\hat\Theta(M_Z)=\left(1+\frac{\cos^2\Theta_W}{\sin^2\Theta_W}
 \Delta\varrho\right) \sin^2\Theta_W
\end{equation}
the correction factors would be different. $r_{X,EW}$, for instance,
would become
\begin{equation}\label{rhat}
\hat r_{X,EW}=1+\frac{x_t}{4 X_0(x_t)}\left(\tau^{(2)}_b+3\right)\xi_t
\end{equation}
This change compensates
for the corresponding change in the coupling constants, which also
differ by terms of ${\cal O}(\xi_t)$ (see (\ref{st2hat})). The
compensation is not exact in our approximation where we use
$X_0(x_t)$ instead of $x_t/8$ as the leading $m_t$-dependent function.
It holds strictly only in the large-$m_t$ limit.
Numerically the different choices for $\sin^2\Theta$ lead to a
difference in the branching ratio by a factor of
$\sin^4\hat\Theta(M_Z)/\sin^4\Theta_W=1.056$
if no higher order electroweak corrections are applied.
After inclusion of ${\cal O}(\xi_t)$ corrections this discrepancy
is reduced to
\begin{equation}\label{s4zw}
\frac{\sin^4\hat\Theta(M_Z)}{\sin^4\Theta_W}\cdot
\frac{r_{X,EW}}{\hat r_{X,EW}}=1.034
\end{equation}
This indicates a reduction of the uncertainty from about
$\pm 2.8\%$ to $\pm 1.7\%$, where these numbers are independent of
the Higgs boson mass.
While $r_{X,EW}$ is smaller than unity by about $1\%$
(see Table \ref{rxytab}) and reduces the larger parameter choice
of $1/\sin^4\Theta_W$, the smaller normalization using
$1/\sin^4\hat\Theta(M_Z)$ is enhanced by roughly the same amount
through $\hat r_{X,EW}$. 
Previous analyses usually employed the latter choice of
$\sin^2\hat\Theta(M_Z)=0.23$, in which case $\hat r_{X,EW}$ is
the appropriate correction factor.

We remark that the ambiguities discussed
here in connection with the weak mixing angle are particularly large
since they are reinforced by a factor of 
$\cos^2\Theta_W/\sin^2\Theta_W\approx 3.5$ as seen in (\ref{st2hat}).
The above estimate of about $\pm 2\%$ for the uncertainty due to as
yet unknown subleading electroweak contributions should therefore be 
quite conservative.

We turn next to a discussion of scheme and scale dependence, which is
useful to further investigate the structure of the electroweak
corrections.
\\
Instead of using the on-shell (pole) definition of $m_t$ (with respect to
electroweak interactions), which we have employed so far, one may
adopt the $\overline{MS}$ scheme for the top quark mass. 
These two
definitions differ by terms of ${\cal O}(\xi_t)$ and are related
by \cite{KS}
\begin{equation}\label{xtbar}
\bar x_t=x_t (1+\Delta_t(\mu,a) \xi_t)
\end{equation}
where $x_t=m^2_t/M^2_W$, $\bar x_t=\bar m^2_t/M^2_W$ and $\bar m_t$
is the $\overline{MS}$-mass. 
The function $\Delta_t$ reads \cite{KS}\footnote{This expression
relates, strictly speaking, the $\overline{MS}$ and the pole 
definition of the top quark Yukawa coupling, which we then write in
terms of the top quark mass.}
\begin{equation}\label{delt}
\Delta_t(\mu, a)=18 \ln\frac{\mu}{m_t}+11-\frac{a}{2}+
\frac{a(a-6)}{2}\ln a+\frac{a-4}{2}\sqrt{a} g(a)
\end{equation}
The corrected Inami-Lim function has been given in (\ref{xxpole})
for the on-shell scheme.
Alternatively we may use the
$\overline{MS}$-scheme (\ref{xtbar}), in which case (\ref{xxpole}) is
replaced by
\begin{equation}\label{xxmsb}
X_0(\bar x_t)+\frac{x_t}{8}\left(\tau^{(2)}_b-\Delta_t + 6 -
\frac{3}{\sin^2\Theta_W}\right)\xi_t
\end{equation}
The ratio of (\ref{xxmsb}) to (\ref{xxpole}), to be denoted by $R$,
provides a measure of scheme dependence (we will put $\mu=m_t$ for
the moment). To linear order in $\xi_t$ we have
\begin{equation}\label{rsr}
R=1+ s_R\ \Delta_t\ \xi_t \qquad
s_R=\frac{1}{X_0(x_t)}
 \left(x_t\frac{\partial X_0}{\partial x_t}-\frac{x_t}{8}\right)
\end{equation}
In the large-$m_t$ limit $X_0\to x_t/8$ and $s_R\equiv 0$, ensuring
the scheme independence of the corrected Inami-Lim function to first
order in $\xi_t$. If we use the full leading order function $X_0(x_t)$,
a residual scheme dependence persists, since the corrections are only
known in the large-$m_t$ limit. Numerically $R=1.002$ for $m_H=300\, GeV$.
This is to be compared with $X_0(\bar x_t)/X_0(x_t)=1.006$, indicating
the scheme dependence when the corrections are altogether omitted.
The scheme dependence is thus reduced from $0.6\%$ to $0.2\%$ in the
decay amplitudes. The effects are twice as big for the branching 
fractions. The scheme ambiguities can be somewhat larger for other values
of the Higgs-boson mass, but the reduction by a factor of three observed 
above is independent of $m_H$. 
\\ 
Note that the reduction in scheme dependence is quite sizable, although
the asymptotic limit is not a good approximation for realistic values
of $m_t$ as $X_0(x_t)\approx 2.8\cdot x_t/8$. This can be understood
by considering the large-$x$ expansion of $X_0(x)$
\begin{equation}\label{xexp}
X_0(x)=\frac{x}{8}+\frac{3 \ln x+3}{8}+\frac{3}{8 x}+
{\cal O}\left(\frac{1}{x^2}\right)
\end{equation}
which shows that the dominant $x$-dependence stems from the leading
term $x/8$.

Related to the scheme ambiguity is the issue of scale dependence,
resulting from the running of the top quark Yukawa coupling. 
This question can be studied using the $\overline{MS}$ formulation
in equation (\ref{xtbar}), which exhibits explicitly the
$\mu$-dependence of $\bar m_t$ due to the Higgs-top Yukawa interaction.
Changing $\mu$ between $100\, GeV$ and $300\, GeV$ results in a 
variation of $X_0(\bar x_t(\mu))$ by $\pm 1.6\%$. This sensitivity is
reduced to $\pm 0.6\%$ when the leading large-$m_t$ corrections
from (\ref{xxmsb}) are included. Unfortunately the residual
$\mu$-dependence in the branching ratios is then still $\pm 1.2\%$,
of the same order of magnitude as the electroweak corrections in
Table \ref{rxytab} themselves.
This indicates again that subleading $m_t$-terms in the electroweak
corrections are important and have to be taken into account if a
higher precision is required.
\\
We finally note that the difference between the two definitions of
$m_t$ in (\ref{xtbar}) amount to typically $1-2\, GeV$. This is still
smaller than the current experimental uncertainty in the top quark mass
of $\pm 5.5\, GeV$ \cite{RAJ}, but will become relevant if future
measurements reduce this error to $\pm 1\, GeV$ or below.
The top quark pole- and $\overline{MS}$-mass in QCD, by contrast,
differ by about $8 GeV$ as already mentioned before. The situation is
similar with respect to the scale dependence. Here the electroweak
scale ambiguity of $\pm 1.6\%$ in the uncorrected lowest order term
$X_0(\bar x_t(\mu))$ may be compared with the corresponding QCD effect
of $\pm 5\%$ ($100\, GeV\leq\mu\leq 300\, GeV$). The latter is reduced
to $\pm 0.5\%$ when the full ${\cal O}(\alpha_s)$ corrections are
included \cite{BB2}.

For definiteness we have restricted our discussion to the function 
$X_0(x_t)$, relevant for $K_L\to\pi^0\nu\bar\nu$ and
$B\to X_s\nu\bar\nu$. Similar observations hold for the decays
$B\to l^+l^-$ governed by $Y_0(x_t)$. 
Here the situation is generally somewhat more favorable, since the
function $Y_0(x_t)$ is closer to its asymptotic limit $x_t/8$
($Y_0(x_t)=1.8\cdot x_t/8$) than it is the case for $X_0(x_t)$.

\section{Conclusions}

In this paper we have investigated the electroweak radiative
corrections of ${\cal O}(G^2_F m^4_t)$ to the decay amplitudes
of $K_L\to\pi^0\nu\bar\nu$, $K^+\to\pi^+\nu\bar\nu$,
$B\to X_s\nu\bar\nu$ and $B\to l^+l^-$.  These corrections arise
at the two-loop level and are the formally leading electroweak
corrections to the one-loop induced FCNC in the limit of large
top Yukawa coupling. Our analysis was motivated by the theoretically
clean nature of the rare decay processes under consideration.
The main benefits of this investigation may be summarized as
follows.
\begin{itemize}
\item
It serves to illustrate the general issues involved in the
calculation of higher order electroweak corrections to rare decays.
\item
It provides a quantitative order of magnitude estimate of these
effects.
\item
It helps to reduce the impact of ambiguities in the definition of 
electroweak parameters on observable quantities.
\end{itemize}

In the large-$m_t$ limit the lowest oder amplitudes are of
${\cal O}(G_F m^2_t)$. The inclusion of the ${\cal O}(G^2_F m^4_t)$
correction eliminates various ambiguities, of order several percent,
that are related to the definition of electroweak parameters in
the lowest order expressions. Such ambiguities exist for instance
between $\sqrt{2}G_F M^2_W$ and $\pi\bar\alpha/\sin^2\Theta_W$ or
between $\sin^2\Theta_W$ and $\sin^2\hat\Theta(M_Z)$, which differ
due to higher order electroweak corrections. Another example is the
uncertainty due to (electroweak) scheme- and scale dependence in
the top quark Yukawa coupling.
\\
Unfortunately the asymptotic, large-$m_t$ limit is not fully realistic
in the cases at hand. Since only the formally leading corrections are
known, the above ambiguities can at present not be removed completely.
They become however smaller when the large-$m_t$ corrections are
applied. Scheme- and scale dependence are reduced by a factor of three
to typically $\pm 1\%$ in the branching ratios. The presumably
largest uncertainty is due to the difference between
$\sin^2\Theta_W=0.224$ and $\sin^2\hat\Theta(M_Z)=0.23$ that leads to a 
change in the lowest order branching fractions by $5.6\%$. At order
${\cal O}(G^2_F m^4_t)$ this is reduced to a total variation of
$3.4\%$. We estimate the uncertainty due to presently unknown
subleading (in $m_t$) electroweak corrections for the top quark
dominated decays $K_L\to\pi^0\nu\bar\nu$, 
$B\to X_s\nu\bar\nu$ and $B\to l^+l^-$ to be about $\pm 2\%$.
In comparison to previous calculations of these rare decays, which
employed $\sin^2\hat\Theta(M_Z)=0.23$ in the overall normalization,
the ${\cal O}(G^2_F m^4_t)$ effects lead to a slight enhancement of
about $1-2\%$ in the central value of the branching ratio.

We remark that the corrections discussed in this paper have no impact
on the extraction of the CKM parameter $\sin 2\beta$ from
$K_L\to\pi^0\nu\bar\nu$ and $K^+\to\pi^+\nu\bar\nu$ \cite{BB6},
as the top contribution essentially cancels out in this case.

An improvement of the uncertainties in the decay rates beyond the
$\pm 2\%$ quoted above would require the explicit calculation
of at least the first subleading two-loop contributions of
${\cal O}(G^2_F m^2_t M^2_W)$. Such corrections are yet unknown for
rare decays, but have been calculated
for the $\varrho$-parameter, relevant for electroweak
precision observables at the $Z$ resonance \cite{DFG}.

In any case our work confirms the expectation that higher order
electroweak effects are well below the experimental sensitivity
in the forseeable future. A further, systematic improvement over the
present situation is however still possible, if it should indeed
appear necessary.


\vfill\eject

\end{document}